TITLE PAGE

# Vulnerability of geriatric patients to biomaterial associated infections: in vitro study of biofilm formation by *Pseudomonas aeruginosa* on orthopedic implants

Authors:

S Dutta Sinha*[1], P.K. Maiti[2], S Tarafdar[1]

[1] Department of Physics, Jadavpur University, Kolkata-700032, India

[2] IPGMER, SSKM Hospital, Department of Microbiology, Kolkata-700020, India

Email: ju.research@gmail.com

Phone number: +91332414-6666 (Office)

Mobile: +91983081510

Fax: +91332433-1078

**Acknowledgements**

The authors are thankful to Department of Science & Technology, Govt. Of India for funding the research through the Women's Scientist Scheme (WOS-A) (Project: DST: SR/WOS-A/LS-466/2013). S Dutta Sinha is thankful to Prof. Dr. Pratip Kundu of School of Tropical Medicine, Kolkata and Dr Debkishore Gupta of Peerless Hospital & BK Roy Research Centre for their helpful suggestions and support.

Abstract

Fibronectin, a glycoprotein secreted by connective tissue cells is found in the human plasma as well as in the ECM. It is known to have an adhesive property and plays a role in cell-to-cell and cell-to-substratum adhesions. In rheumatoid arthritis (RA) and osteoarthritis (OA), fibronectin is locally synthesized by the synovial cells, and the synovial fluid level of fibronectin is found to be double that in plasma. The concentration of fibronectin in the synovial fluid under such conditions is found escalate to 2-3 times higher than in the corresponding plasma. The present article demonstrates that the protein has a strong tendency to get adsorbed on biomaterials after an implant surgery in preference to lighter proteins like albumin, which in turn enhances the growth of robust biofilms. The present article demonstrates that, heightened risk of dangerous implant infections due to the formation of such biofilms coupled with the degrading immunity levels of the geriatric patients have the potential to transform an implant surgery to a 'life threatening' event.

**Introduction**

Degenerative Arthritis is a well-known cause of morbidity of ageing population (1-3) and has got a large spectrum of problems and functional limitations (4-6). It starts in the third decade of life and usually manifests in those areas of human bony structure through which the line of weight-bearing passes (7-8), from the Bio-mechanical point of view. The lowermost of such areas is the knee, where the degenerative process involves the cartilage structure, which erodes gradually with course of time (as an individual grows older) through day-to-day activities and life-style, in an irregular pattern, depending on the variety of use of the joint (9-12). An inflammation cascade sets in involving internal tissues predominantly the synovial membrane, resulting in synovitis (13-14). Pain becomes the most obvious symptom and gradually becomes a determinant of discomfort in performing daily activities. Patients need various combinations of medicine and lifestyle modifications to maintain a relatively pain-free mobility. However, with advancing age and due to individual variation of pattern of pathological progression up to critical amount of cartilage loss and adjoining bone erosion, certain percentage of affected population experience significant deterioration of daily activities due to pain and stiffness (15-16).It is at this juncture, patients resort to Replacement surgery (17-20), where the cartilage and adjoining bone portion is removed and replaced by implant. At this stage, medical implants serve as a source of relief to a large section of population especially in the geriatric age group, where implants may be required to replace the hip joints, knee joints etc.

Biomaterials (21-22) which should be used for successfully designing medical implants, have to face a number of challenges. First, all biomaterials must be biocompatible (23-24) and, unless the material is meant to degrade within the body, it must offer long-term resistance to biological attack in vivo. Biomaterials which are prone to evoke cellular interactions in vivo are liable to become bioactive or trigger an inflammatory response (25-26) after introduction in a physiological environment. Load-bearing devices such as orthopedic implants, in addition face the challenge of a coupled effect between the structural requirements of the implant and the aggressive environment of the body.

Polymers remain the most versatile class of biomaterials (27), which have been extensively applied in the manufacture of implants, because they offer the unique benefit of being intrinsically resistant to environmental attack. The two main challenges of extended use of implanted biomaterials are the possibility of biomaterial–centered infection and lack of successful tissue integration. These seemingly disconnected phenomena are actually similar expressions of cell-substratum surface interactions. In this article we have chosen the polymer high density polyethylene (HDPE) for our discussion. It is extensively used in orthopedic implants and hence has an immense impact on the well being of the elderly after a knee or hip replacement surgery. Total knee or hip arthroplasty typically employs an HDPE insert that articulates against a cobalt-chromium alloy or ceramic in order to restore function to a damaged or diseased knee or hip joint.

Implants have an inherent tendency to be coated by host proteins from the plasma such as fibrinogen and fibronectin shortly after implantation (28). Initially, fibrinogen/fibrin seems to be the dominant-coating host protein, while fibronectin becomes dominant in the long term, since fibrinogen/fibrin is degraded in course of time. Fibronectin is a large glycoprotein found in body fluids (29), on the surfaces of cells and in the extracellular matrix (ECM). There are mainly two major forms of fibronectin: plasma fibronectin and cellular fibronectin. Incidentally, during acute degenerative arthritis fibronectin gets concentrated in the synovial fluid (30-31) and cells lining the synovial membrane since fibronectin has relation with inflammation and wounds (32-33). It has been detected that two predominant generic species of FN migrating under reducing conditions at ~200+ and ~170 kd, respectively, within SF samples from patients with either OA or RA, regardless of whether samples were obtained via needle aspiration or arthroscopic lavage. Subsequent knee or hip arthroplasty involves removal of the damaged joint but removal of fibronectin from the neighbouring tissue is practically an impossible task. Hence the introduction of an implant into such an aggressive physiological environment enhances its probability of being adsorbed with a robust layer of fibronectin, not only from the plasma or ECM but also from the tissue and cells neighbouring the previously diseased joint. Implants can then act as a colonization surface to which many bacterial species readily adhere (25,34).

We have previously demonstrated that, the amount of adsorption of protein (BSA) on biomaterials can affect the nature and quality of biofilms formed on them (35). Our present endeavor is to demonstrate the effect of adsorption of fibronectin (Fn) on high density polyethylene (HDPE) in vitro, and precisely analyze the biofilm architecture of a gram negative nosocomial pathogen Pseudomonas aeruginosa on it. The in vitro model with Fn adsorbed on HDPE in a random configuration gives an oversimplified view of the in vivo environment of introduction of foreign bodies within a physiological environment. The single protein adsorption model used here is meant to understand in detail the manner in which Fn affects the biofilm formation by a gram negative bacterium on immuno-compromised patients. Artificial blood or normal human blood, having a number of proteinaceous and non-proteinaceous components was not used in our study, as a multi-component adsorbate would make the study not only cumbersome but also prevent the understanding of the contribution of a single protein in implant infections. This understanding is significant, taking into consideration the relevant facts that, HDPE is widely used in orthopedic implants, Fn is related to inflammation and infection and orthopedic replacement surgeries are frequently performed on the geriatric age group.

## 2. Materials and methods

**2.1 Biomaterials and production of the conditioning layer.**

Commercially available HDPE used in orthopedic implants was obtained after fine machine polishing in square configuration (10mmx10mm) from Plastic Abhiyanta Ltd, India, which we have referred to in the article as HDPE chips. After polishing, the samples were cleaned by 2 min ultrasonication in a 35 kHz ultrasonic bath (Rivotek Instruments, India) and thoroughly rinsed with demineralized water. The water used in all our experiments was of HPLC grade (Lichrosolv) from Merck, India. Tris EDTA buffer solution pH 7.4 was obtained from Sigma Aldrich, USA and Fibronectin from human plasma (CAS Number 86088-83-7, MDL Number MFCD00131062) lyophilized powder, 45 kDa and Bovine serum albumin were obtained from M P Biomedicals, USA.

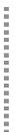

Since HDPE is not autoclavable, the chips were rinsed twice with ethanol, blow dried and preserved in a vacuum desiccator. A section of the experimental chips were immersed in reconstituted Fibronectin solution of concentration 1 microgm/l for 24 h. After 24 h, all chips with Fn- conditioning flayer were dipped three times in demineralized water. A small section of the untreated, cleansed chips were coated similarly with BSA by immersing them in reconstituted BSA solution of concentration 1.5 gm/ml for 24 h and rinsed thoroughly with demineralized water. The adsorption time 24 hours is also known as the exposure time and will henceforth be referred to as $\tau$. Rest of the chips were left untreated to serve as control. The concentrations of proteins were kept in proportion with their respective concentrations in human blood. The chips were all treated at 30 deg C, and $\tau$ was kept sufficiently high to obtain a complete surface coverage. After the production of the respective conditioning layers, the treated chips were thoroughly rinsed in PBS in order to remove any non-adsorbing proteins, finally leaving only irreversibly adsorbed proteins on the HDPE surface. The untreated chips were also rinsed with demineralized water and PBS, and all the experimental chips were finally blow dried and preserved in a vacuum desiccator.

**2.2 Bacterial strain and culture condition**

A number (25) of bacterial strains were clinically isolated from uro-catheters of patients having urinary tract infections from the Department of Urology, Institute of Post Graduate Medical Education & Research, Kolkata with a six month period. The bacterial strain *Pseudomonas aeruginosa* UC10 used in this study was reported to be the strongest biofilm former among all the clinically isolated strains by test tube biofilm assay [36] performed in the Department of Microbiology, Institute of Post Graduate Medical Education & Research. We have also used a reference strain of *Pseudomonas aeruginosa*, ATCC 27853, for producing biofilms on exactly the same *substrates,* under similar conditions to compare the architecture of biofilms produced by the clinical strain and the reference strain.

Each of the above type of frozen culture was initially inoculated on Tripticase soy agar and incubated at 37 deg C for 24 hours. Each culture was then transferred via swab to a buffer solution and a suspension equivalent to a Macfarland 0.5 (~1.5 x $10^8$ CFU/ml) was prepared. This suspension was diluted 1:100 and 1 ml was used to inoculate 100 ml of sterile LB broth [Luria Bertanii Agar (LBB) obtained from Himedia, India].The bacteria grown overnight in LBB at 37° C were diluted in the same medium to an optical density of 0.5 at 600 nm and ready for use as bacterial culture for growing biofilms.

**2.3 Growing of biofilms and analysis**

Biofilms were grown on both treated and untreated HDPE chips with the clinical strain and reference strain by the following method:

The diluted culture of the bacterial strain obtained in **Section 2.2** was poured over the surface of the treated chips obtained from **Section 2.1** placed in the wells of 24 well tissue culture plates (Tarsons, India) and on untreated chips kept in a separate 24 well plate. The sets of chips having BSA and Fn adsorbed on them were also kept in separate 24 well plates and each system was closed and sealed with paraffin without addition or removal of any component with the exception of broth. The sterile LBB was added carefully from time to time to avoid desiccation and incubated at 37°C for 7 days with shaking at 180 rpm. Each set of experiment was performed in triplicate. The plates were sealed and placed on the shaker plate of the BOD incubator set at 180 rpm and rotated simultaneously for the entire growth period. All chips were incubated for 7 days, except a few untreated chips which were kept for FTIR analysis, were incubated for a 14 day period.

After the entire growth period, the polymer chips were aseptically removed and washed with phosphate buffered saline (PBS pH 7.2). This step eliminated all the free floating bacteria and only the sessile forms remained attached to the surface, after which the chips were air dried and preserved in a vacuum desiccator.

**Preparation for FESEM analysis and FTIR measurements**

To compare the architecture of the biofilms produced by the clinical strain and the reference strain of *P aeruginosa* after 7 days, FESEM measurements were conducted at 5.0 kV-10 kV in a field emission scanning electron microscope (FESEM: Inspect F50, FEI Europe BV, and The Netherlands; FP 2031/12, SE Detector R580).

The chips with attached bacterial cells were covered with 2.5% glutaraldehyde and kept for 3hours in 4°C after which they were washed thrice with the phosphate buffer solution. They were then passed once through the graded series of 25, 50 and 75 % ethanol, twice through 100% ethanol each for ten minutes, finally transferred to the critical point drier and kept overnight to make them ready for biofilm analysis.

A section of the above chips with biofilm of the reference strain had either BSA or Fn as the conditioning layer. A small section of such chips were kept separately and directly used for FTIR analysis. An FT-IR spectrophotometer from Shimadzu equipped with a liquid-N2- cooled, medium-range mercury-cadmium-telluride detector (4,000 to 400 cm-') was used to collect the FT-IR spectra of the polymer chips with biofilms of the reference strain.

Rest of the treated and untreated chips with biofilm of either strain, were used for FESEM measurements to analyze the difference in the effect of Fn adsorption on the biofilm architecture.

## 3. Results

The scanning electron micrographs of biofilms of clinical strain and reference strain of *P aeruginosa* on bare HDPE surface and the same surface coated with Fn were obtained by applying potentials ranging from 5 kV to 17 kV, while EDAX measurements required the application of a maximum potential of 70 kV.

### 3.1 Biofilms on bare HDPE surface

The 7-day old biofilms produced by the clinical strain of P aeruginosa on bare HDPE surface has a thin sparse exopolysaccharide matrix as shown in Fig 1a, while the biofilm of reference strain on the same surface is uneven and has the topography of an 'ant-hill' (Fig 1b). The presence of newly dispersed bacteria scattered on the surface of the biofilm is vivid for the clinical strain, but bacteria of the reference strain remain concealed within the exopolysaccharide encasing of its biofilm, and is revealed only when a high potential is applied on the sample at that point during EDAX measurements.

### 3.2 FESEM analysis of biofilms on HDPE

The bacteria of clinical strain forms a smooth and robust biofilm on HDPE surface adsorbed with Fn as shown in Fig 2A, proving the presence of unique cell adhesion property of Fibronectin, which is displayed even though the protein is liable to get denatured after adsorption on HDPE. The biofilm has pores and channels for gaseous exchange and nutrient supply to the cells residing inside the biofilm. In. case of the reference strain, the biofilm formed (as shown in Fig 2b), is thick and fluffy in appearance but differs from the former in surface topography. Pores and channels are not visible in the ultra-structure but unusual presence of some crystals on the biofilm surface is a mysterious finding. EDAX measurements to identify the composition of such crystals (Fig 3) reveal the presence of NaCl, though the source of its production is yet unknown.

Fig 4a and 4b demonstrates crack formation in biofilm of the clinical strain on HDPE surface adsorbed with Fibronectin and BSA respectively. Wedge-shaped crack in the biofilm (in Fig 4a) unleashes the strong adhesive nature of the substrate constituting Fibronectin and HDPE. Cracks usually open up in a film due to desiccation or mechanical stress (37-38) arising within it. In this instance a simultaneous action of a strong adhesive force of the substrate and robustness of the biofilm might have lead to the opening of a wedge shaped crack in the biofilm (39). The strong adhesive force may be linked also to the contribution of the presence of fibronectin binding proteins on the cell surface of P aeruginosa. The thick and virulent biofilm on the fibronectin covered surface as shown in Fig 4a if produced in vivo, is liable to cause loosening of the implant, preventing tissue integration which may finally result in an implant failure. Such a cascade of events will definitely incur serious detrimental effects on the physiological and psychological well being of elderly patients who are devastated by the above sequence of events. If the course of events turns far worse, such patients may be victims of depression and irrecoverable septicemia [40] which may even lead to death. Fig 4b shows cracks which reveal a clean surface beneath the thick biofilm which proves that the polymers present within the biofilm have a strong cohesive property and the cohesive force is greater than the adhesive force between the substrate (composed of HDPE and BSA) and the biofilm. In such an instance, the biofilm on becoming robust might have the probability of getting dislodged as a whole and clogging the blood vessels causing thrombosis.

### 3.3 FTIR analysis of biofilms on HDPE

FTIR results in Fig 6 compares the FTIR data obtained for 7-day old biofilms by the reference strain on HDPE adsorbed with BSA and Fn, subtracting the contribution of the polymer (HDPE), which is identical in both the cases. These results can be summed up considering the following spectral windows:

i) The window between 3000 and 2800 $cm^{-1}$ ( the 'fatty acid region' **I),** dominated by the -CH3, >CH, and =CH stretching vibrations of the functional groups usually present in the fatty acid components of the various membrane amphiphiles.

This component is present in the biofilm on BSA covered HDPE and totally absent in Fn covered HDPE. Hence biofilm on BSA covered HDPE has significantly proportion of fatty acids, while biofilm on the Fn covered surface is completely devoid of it.

ii) The window between 2200 and 2500 $cm^{-1}$ (no special designation of any particular functional group)

Sharp peak present around 2350 $cm^{-1}$ for biofilm on Fn covered HDPE, but no peak for BSA covered HDPE. The sharp peak is due to the presence of environmental $CO_2$ and this feature is not significant for our analysis.

iii) The window between 1800 and 1500 cm$^{-1}$ ( the 'amide region'), dominated by the amide I and amide I1 bands of proteins and

peptides ;

Peaks are present around 1875 cm$^{-1}$, for Fn covered HDPE and not for BSA covered HDPE proving the presence of proteins and peptides in the biofilm on Fn adsorbed surface and not in the one adsorbed with BSA. The proteins and peptides in this frequency range are therefore characteristic of the EPS (exopolysaccharide) matrix of the biofilm on the Fn coated HDPE, and not a contribution of either adsorbate, as they do not correspond to either Fn or BSA.

iv) The window between 1500 and 1200 cm$^{-1}$ ( the 'mixed region'), a spectral region containing information from proteins, fatty

acids and phosphate-carrying compounds ;

Both Fn and BSA covered HDPE surfaces have sharp and relatively much higher peaks in this region, proving the presence of proteins, fatty acids and phosphate carrying compounds. However it is quite clear that the nature of components in the BSA covered and Fn covered HDPE surfaces are distinctly different. This region is liable have some contribution from either adsorbate.

v) The window between 1500 and 1400 cm$^{-1}$ ( the 'fatty acid region II'), a subrange of the previous window dominated by the -CH3 and >CH2 bending vibrations of the same functional groups.

Sharp peaks present for Fn covered HDPE surface at around 1450 cm$^{-1}$ showing the stark presence of CH$_2$- bending lipids and one peak for BSA covered HDPE surface around 1450 cm$^{-1}$ is much lower in intensity.

vi) The window between 1100 and 1350 cm$^{-1}$ ('nucleic acid' region) a subrange of the spectral window in (iv), showing the presence of DNA/RNA.

Peaks in this region have much higher intensity for Fn covered HDPE showing the presence of more nucleic acid components in the biofilm on it compared to that on BSA covered HDPE. This feature can be directly utilized in quantifying and comparing the biofilm activity on the two surfaces. As a consequence it can be concluded that a biofilm of similar age on a Fn covered HDPE surface is prone to have more pathogenic bacteria than that on a BSA covered one having the same dimensions and identical $\tau$. Since amount of pathogenicities can be directly related to virulence factors for a pathogenic bacteria, we can conclude that an HDPE implant adsorbed with Fn will be more prone to infection compared to the one adsorbed with BSA.

vii) The window between 1200 and 900 cm$^{-1}$ ( the 'polysaccharide region'), dominated by the fingerprint-like absorption bands of the carbohydrates present within the cell wall of the bacteria.

The peaks for the Fn covered and BSA covered HDPE are not at all synchronized or of the identical intensity.

The above feature proves unambiguously, that the cell wall of the biofilm bacteria (both P aeruginosa ATCC 27853) changes its composition when encountering different substrates.

vii) The window between 900 and 450 cm$^{-1}$ ( the 'true fingerprint') showing some remarkably specific spectral patterns, are as yet unassigned to cellular components or to functional groups.

**Discussion**

Our experiments in vitro reveal that *P aeruginosa* a dangerous nosocomial pathogen, normally affecting immuno-compromised patients suffering from cystic fibrosis (41-42), can be a tremendous threat to patients undergoing replacement surgery provided the pathogen gets access to the implant surface. On the basis of the FTIR results it can be concluded that biofilm formed on Fn adsorbed orthopedic implants can be a potential source of post-operative implant infections. This is a major point of concern to patients of all age groups as biofilm associated infections are antibiotic resistant. In addition, patients undergoing replacement surgery being mostly in the age range > 65 years when the immune system gradually becomes less effective (43-44), tend to be most vulnerable to such infections.

Also as people age, which is an independent risk factor for postoperative complications and mortality (45), macrophages (which ingest bacteria) destroy bacteria more slowly (46). This slowdown may be one reason that microbial infections affect older people more seriously. T cells respond less quickly to the antigens (47-48) and there are fewer white blood cells capable of responding to new antigens. Thus, when older people encounter a new antigen, the body is less able to remember and defend against it (49). Older people have smaller amounts of complement proteins and do not produce as many of these proteins as younger people do in response to bacterial infections. The amount of antibody produced in response to an antigen is less, and the antibodies are less able to attach to the antigen. All these above physiological changes may explain why implant associated infections are not only more common among older people (44, 47) but may even turn fatal in absence of adequate and timely treatment.

All the above factors coupled with our in vitro experimental results reveal that the sterilization of the device to be implanted is not the only factor that should be taken care of during an implant surgery, but care should also be taken that a post operative haematogenous seeding does not occur from auxiliary devices such as catheters. Keeping in mind that changes in immune function of the older people may contribute to their greater susceptibility to infections, special care should be taken for their protection and safety not only by the surgeon but also by the assisting healthcare staff. In addition certain unavoidable situations, such as adsorption of fibronectin on the implant surface from the cells and tissues neighbouring a previously diseased arthritic joint or from remnants of drained out synovial fluid may serve as a potential ingredient for future biofilm formation and subsequent implant infections. Hence apart from the precautions of the hospital staff and surgeons towards peri-operative and postoperative infections, an alteration of the surface of the biomaterial might prove to be of great advantage in deterring implant infections. Adhesion to HDPE, (a hydrophobic, negatively charged surface) has been shown to be influenced both by surface hydrophobicity and electrophoretic mobility and can be best understood in terms of DLVO theory (50). Our experiments reveal that a coating of fibronectin on HDPE, coupled with the presence of Fibronectin binding proteins in the gram negative bacteria *P aeruginosa* (51) poses the proneness of biofilm associated infections after an implant surgery. The modulation of surface characteristics of HDPE, so that it is simultaneously successful in inhibiting fibronectin adsorption and promotion of tissue integration might be a source of great relief to patients belonging to the geriatric age group, since they are worst affected by implant infections.

The innovation of smart biomarkers (52) for detecting an implant infection at an early stage might also prove to be of immense help and provide enough scope of suitable treatment. The nano-pores and nano-channels present in the biofilm (Fig 2 a) might provide an opportunity to use nano-medicines (53-54), for treating biofilm associated infections effectively, which are other';l'wise resistant to antibiotic therapy. Treatment of biofilm associated infections using nano-medicines may open a new vista to efficacious biofilm therapies in the case of geriatric patients having low immunity levels. In case such innovative methods come into practice, it is the elderly generation which will benefit most from the evolution in the treatment of implant associated infections may be even at advanced stages. An Enhanced recovery protocol for hip and knee replacement (55-56) has been suggested, which is supposed to reduce post-operative deaths after implant surgery especially in the elderly patients. It assumes that multimodal intervention may reduce stress-induced organ dysfunction and the accompanying morbidity, that results in the subsequent need for hospitalization (57-58). It was found that there was a substantial decrease in mortality and some early complications following initiation of the enhanced recovery protocol in unselected and consecutive series of 4,500 primary joint replacements performed by the same group of surgeons.

**Conclusion**

We conclude that, the present study reports for the first time that adsorption of fibronectin (Fn) on an orthopedic implant surface, may be a potential source of post-operative implant infection, which may even turn life endangering for the geriatric patients undergoing orthopedic replacement surgeries. The above finding, coupled with the multiple factors plaguing the immunity level of elderly patients suggests that, it is significant to have a correct understanding of the manner in which a collaborative process across disciplines and partnering with hospital departments can be developed to study, monitor and care especially for elderly patients undergoing orthopedic replacements.

**Funding**

This project was funded by the Department of Science & Technology, Govt. Of India for funding the research through the Women's Scientist Scheme (WOS-A) (Project: DST: SR/WOS-A/LS-466/2013).

**Acknowledgements**

S Dutta Sinha is thankful to Prof. Dr. Pratip Kundu of School of Tropical Medicine, Kolkata and

Dr Debkishore Gupta of Peerless Hospital & Research Centre for their helpful suggestions and support.

**FIGURE CAPTIONS**

Fig 1. Biofilm on bare HDPE surface by (a) clinically isolated of strain *P aeruginosa* (b) reference strain *P aeruginosa* ATCC 27853.

Fig 2. Biofilm on HDPE adsorbed with Fn by (a) clinically isolated strain of *P aeruginosa*, inset shows magnified view of the biofilm with pores and channels (b) reference strain *P aeruginosa* ATCC 27853

Fig 3. EDAX of biofilm on Fn coated HDPE by (a) clinically isolated strain of P aeruginosa (b) reference strain P aeruginosa ATCC 27853

Fig 4. Crack in biofilm of clinical isolate of *P aeruginosa* on (a)HDPE adsorbed with fibronectin

(b) HDPE adsorbed with BSA

Fig 5 Development of fimbriae on the surface freshly dispersed bacteria in a biofilm of clinically isolated P aeruginosa.

Fig 6. Comparison of FTIR data of 7-day old biofilms of *P aeruginosa* ATCC 27853 formed on HDPE surface adsorbed with BSA and Fn

**Compliance with Ethical Standards**

**Disclosure of potential conflicts of interest**

**Funding:** This study was funded by Department of Science & Technology, Govt. Of India through the Women's Scientist Scheme (WOS-A) (Project: DST: SR/WOS-A/LS-466/2012)

**Conflict of Interest**: The authors declare that they have no conflict of interest.

**Research involving Human Participants and/or Animals:** No research was carried out with human participants/or animals.